# Dislocations and Grain Boundaries in Two-Dimensional Boron Nitride

*Yuanyue Liu, Xiaolong Zou, and Boris I. Yakobson\**

Department of Mechanical Engineering and Materials Science, Department of Chemistry, and the Smalley Institute for Nanoscale Science and Technology, Rice University, Houston, Texas, 77005, USA.

\* To whom the correspondence should be addressed. Email: biy@rice.edu

**ABSTRACT:** A new dislocation structure—square-octagon pair (4|8) is discovered in two-dimensional boron nitride (h-BN), *via* first-principles calculations. It has lower energy than corresponding pentagon-heptagon pairs (5|7), which contain unfavorable homo-elemental bonds. Based on the structures of dislocations, grain boundaries (GB) in BN are investigated. Depending on the tilt angle of grains, GB can be either polar (B-rich or N-rich), constituted by 5|7s, or un-polar, composed of 4|8s. The polar GBs carry net charges, positive at B-rich and negative at N-rich ones. In contrast to GBs in graphene which generally impede the electronic transport, polar GBs have smaller bandgap compared to perfect BN, which may suggest interesting electronic and optic applications.

Dislocations and grain boundaries (GB) play an important role in the properties of materials.[1] Dislocations in carbon hexagonal lattice have been well documented,[2-6] and recently significant progress has been made to reveal the structures of GBs[7-13] and their influence on properties of graphene.[9,14-20] However, little[21,22] is known about dislocations and GBs for 'white graphene': two-dimensional (2D) hexagonal boron nitride (h-BN), in spite of its promising application in nano-electronics and optoelectronics.[23] Although both materials are one atom thick, mechanically strong against stretching and flexible under bending,[23] the hetero-elemental nature of h-BN brings more complexity compared to homo-elemental graphene. In graphene, the energy of dislocation is dominated by its elastic strain, therefore the core of dislocations is constituted by pentagon-heptagon pairs (5|7s) because of their lower strain energy than other polygon pairs, for example, square-octagon pairs (4|8s). In BN, the energy of dislocation consists not only of topological stain but also possible homo-elemental bonding, which is weaker than hetero-elemental bonding in perfect lattice. 5|7s have lower strain energy, but they inevitably introduce homo-elemental bonding (either B-B or N-N); 4|8s are free of any homo-elemental bonding while suffering from higher strain energy. Which one constitutes the core of dislocation? The delicate balance between the strain and chemical bonding contributions calls for quantitative analysis, as presented in this work. Knowing the structure of dislocation, we are able to further study structures and properties of GBs, which are composed of aligned dislocations. In the synthesis of BN by chemical vapor deposition (CVD),[24] GBs are formed where two grains meet each other by propagating growth fronts. The growth fronts can have elemental polarity, for example, B (or N) terminated edges, which may lead to B-rich (or N-rich) GB. The growth fronts can also retain B:N=1:1 as interior of materials, which may lead to non-polar GB. Although overall polycrystalline BN keeps its stoichiometry as 1:1, it can locally deviate from this balance along GB. Would those polar GBs 'react' with each other by pairing of their extra atoms through diffusion and eventually become non-polar? Or would non-polar GBs spontaneously develop polarity allowing enough diffusion time? How do these behaviors correlate with the structure of GBs? Under local thermodynamic equilibrium, these questions can be answered by the energies of GBs. Finally, the properties of GB are discussed, featuring charge accumulation and bandgap decreasing.

**RESULTS/DISCUSSION:**

We start our discussion by searching for structure of dislocation core. Dislocation with burgers vector $\boldsymbol{b} = (1,0)a$, where $\boldsymbol{a}$ is the lattice vector of 2D h-BN, corresponds to removal of one armchair (AC) atomic chain from perfect lattice. Its core is constituted by a single 5|7. However, dislocation with larger burgers vector $\boldsymbol{b} = (1,1)a$, corresponds to removal of two zigzag (ZZ) chains from perfect lattice.[21] Its core can be constituted by a single 4|8, as shown in Figure 1a, or alternatively, by two 5|7s with homo-elemental bonds, obeying the equation $(1,1) = (1,0) + (0,1)$. These 5|7s can be thought of as being formed by insertion or subtraction of BN dimers sequentially to dislocation core(s). Our calculations indicate that 4|8 is more energetically favorable than 5|7s, by 1.05, 1.51, and 2.29 eV, from left to right in Figure 1a. Figure 1 shows only one set of 5|7s which have N-N bond spatially close to the compressed region of the sheet and B-B close to the expanded part (denoted as N-N|B-B). The other set of 5|7s (B-B|N-N) is also energetically unfavorable than 4|8 by 1.84, 0.94, and 2.12 eV, respectively. These results suggest 4|8 as possible dislocation core of BN. In contrast, the same type of dislocation in graphene is constituted by 5|7s which are adjacent to each other according to our calculations. This indicates that BN is able to avoid the unfavorable homo-elemental bonding at the cost of increasing lattice strain. In fact, the lattice strain in 2D system is eased by relaxation in the third dimension.[7,11] All of these defects induce buckling of free standing sheet in the out-of-plane direction (Figure 1b), which effectively relaxes the strain from dislocation core. Constraining the sheet to be planar hinders strain relaxation and reverses the energy order of 4|8 and 5|7s: 5|7s energy is down from the 4|8 to -2.43, -5.61, -6.96 eV for N-N|B-B, and -0.09, -5.42, -6.97 eV for B-B|N-N. Clearly, the preference of 4|8 is facilitated by buckling, which is one distinct feature of co-dimensional 2D systems.[25] However, it should be noted that the dislocation dipole, formed under mechanical tension, is composed of two 5|7s, instead of 4|8s.[21] This is because, if written in terms of dislocations, 5|7s are generated by (0,0) = (1,0) + (-1,0), while 4|8s are obtained by (0,0) = (1,1) + (-1,-1). The larger Burgers vectors $|b|$ in 4|8 dipole make it less favorable than 5|7s.

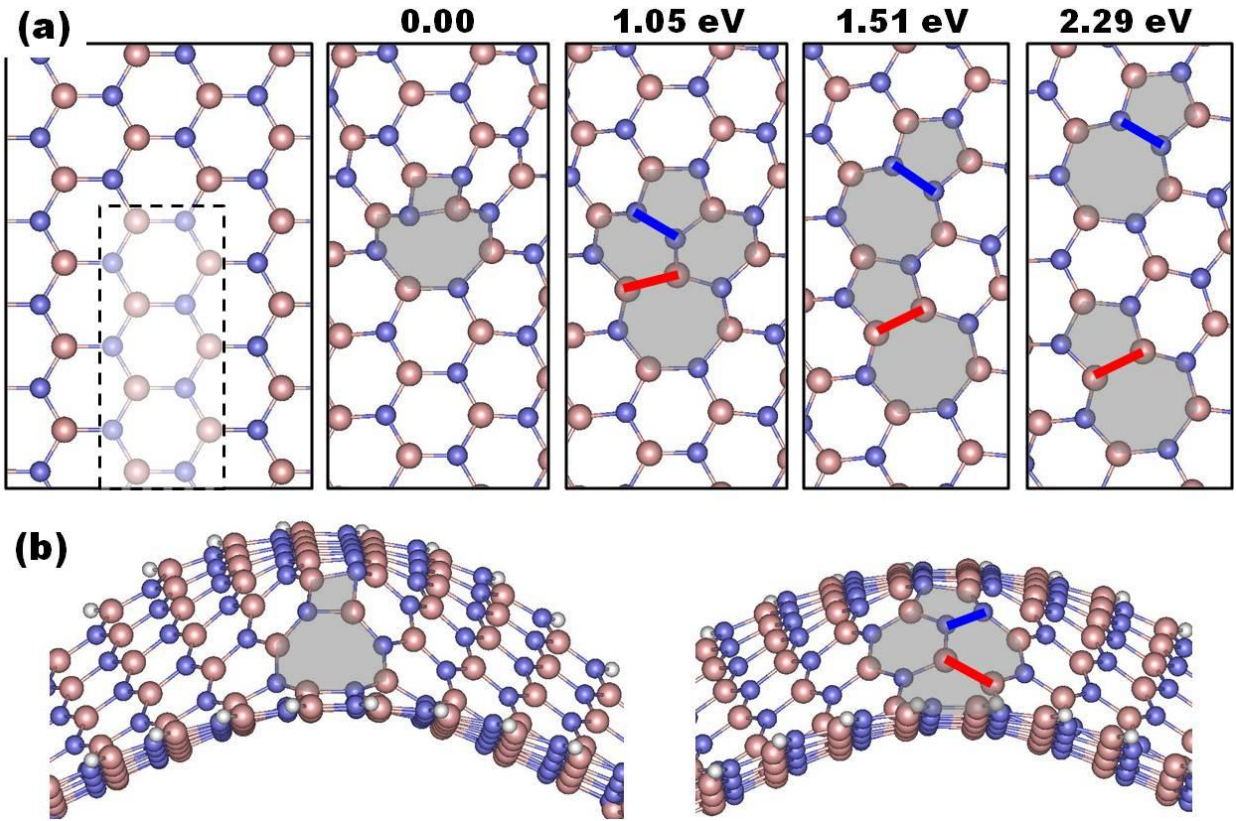

**Figure 1**. (a) Top view of dislocation core with burgers vector $|b|=\sqrt{3}a$, where $a$ is the lattice parameter of 2D h-BN. From left to right: two ZZ lines are cut from hexagonal lattice, dangling

bonds are reconnected forming 4|8, 5|6|7, adjacent 5|7s or sparse 5|7s. (b) Side view of the dislocation constituted by 4|8 (left) or 5|6|7 (right). Red balls represent B and blue ones for N. Red (blue) lines highlight B-B (N-N) bonds. Dislocation cores are marked by shadows.

Having known the structures of dislocations, we can construct the GBs by aligning dislocations with their density prescribed by the tilt angle between two titled grains, $\alpha$. Here we focus on GBs which bisect $\alpha$ ($\alpha = 0°$ means two grains connected along AC direction. The illustration of $\alpha$ is shown in Supporting Information.) since they have lower energies compared to inclined GBs. Different from GBs in graphene, where $\alpha$ has only $120°$ periodicity, two grains of BN have to rotate by $240°$ in order to recover their initial positions, as shown in Figure 2. However, $\alpha$ is not sufficient to specify all the GBs. There are two distinct families of GBs which have to involve another variable to distinguish them: (i) grains in the first family (right side of Figure 2) have mirror symmetry with respect to GB, thus called symmetric GB (sym-GB) later. (ii) In the asymmetric GB family (asym-GB, left side of Figure 2), one grain can be viewed as mirror operation of the other one plus additional inversion. (These two families can be transformed into each other by planar rotations of individual grains; however, due to the imposition of constraint that GBs bisect $\alpha$, this distinction between *sym* and *asym* became necessary.)

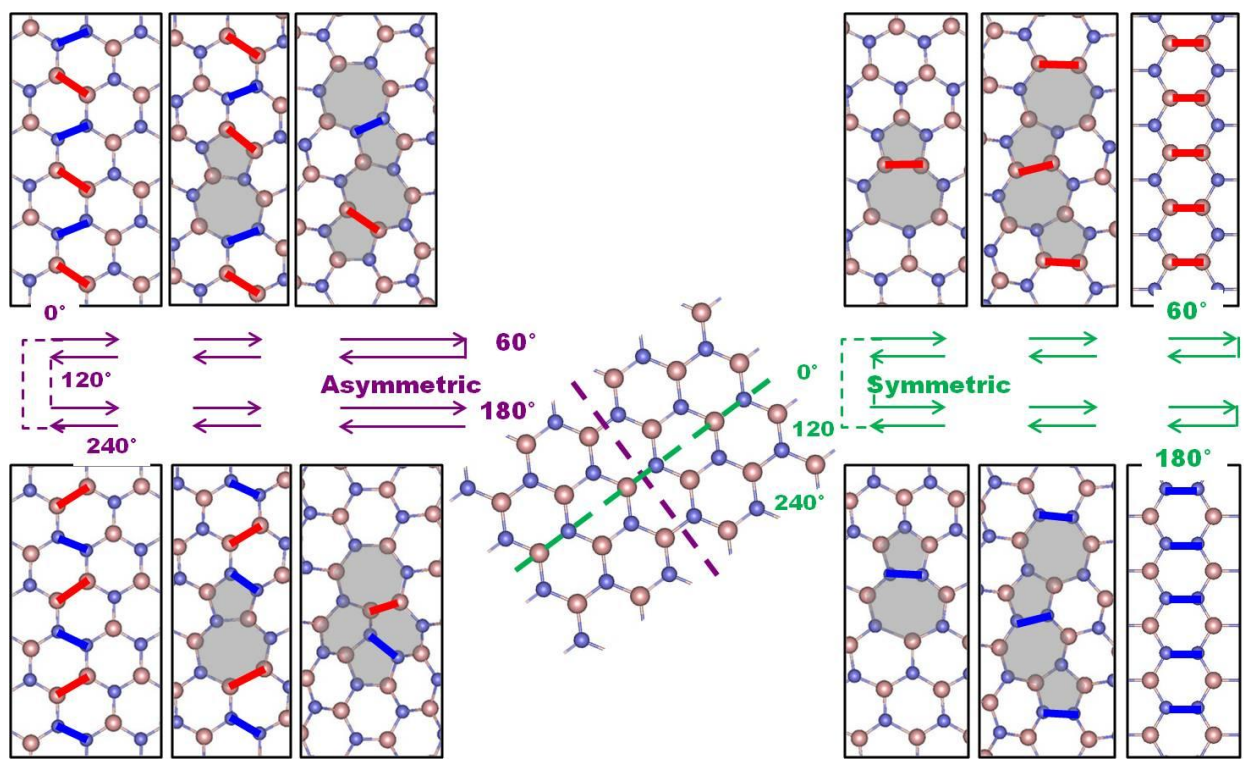

**Figure 2**. Structures of GBs constituted by 5|7s as a function of tilt angle. The middle panel shows a perfect BN lattice which can be thought as two grains are connected perfectly either along green or purple dashed lines. Rotation of two grains with respect to green dashed line leads to symmetric GBs (right panel), and purple dashed lines result in asymmetric GBs (left panel). See details in the text.

Asymmetric GBs can be possibly constituted by 5|7s, similar to graphene. Figure 2 shows their atomic structures. Following the purple arrows, GBs transform from AC junctions, to straight 5|7s, to 5|6|7s (or inclined 5|7s depending on energy preference), then to ZZ junctions, and full way back to start the next 120° circle. 60° and 180° GBs correspond to perfect BN, while 0°, 120°, and 240° ones have alternating B-B and N-N bonds along GB. For all the GBs, there are always equal amount of B-B and N-N bonds, which can be fully eliminated by reconstruction to 4|8s (left side of Figure 5). Their energies are shown in Figure 3 by purple scatters (squares for 4|8s and circles for un-reconstructed 5|7s) as examples. It can be clearly seen

that 4|8s are always lower in energy, even at 0⁰ (60⁰ and 180⁰) where GBs are made by hexagons. Although sparse dislocations induce the buckling in the direction perpendicular to the sheet, increase of defects concentration reduces buckling due to the cancellation of strain field between two neighboring dislocations. For asym-0⁰ (120⁰ and 240⁰), the planarity is retained. The buckling of GB can be measured by its inflection angle, as shown in Supporting Information. In short, the lower energy asym-GBs are constituted by 4|8s.

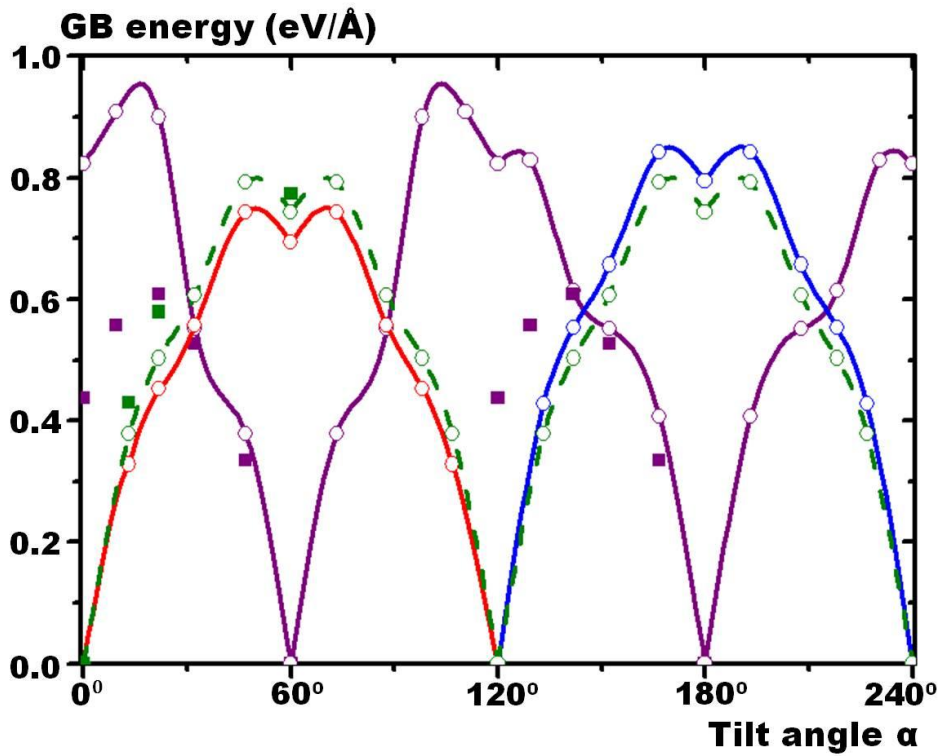

**Figure 3**. Energy of GB as a function of tilt angle $\alpha$. Scatters are computed data, which are connected by lines. Hollow circles are GBs made by 5|7s, and solid squares are for 4|8s. Purple for asym-GBs, red for B-rich sym-GBs, blue for N-rich sym-GBs, and green for the average energy of B-rich GBs and their corresponding N-rich analogs. Note that energies of polar GBs depend on chemical potential, thus red and blue lines are shown just for illustration.

Sym-GBs can also be possibly constituted by 5|7s, but with elemental polarity (either B- or N-rich) along GB except several special $α$ where perfect BN is recovered. Following the green arrows in Figure 2, GBs transform from AC junctions, to B-rich straight 5|7s, to B-rich inclined 5|7s, then to B-rich ZZ junctions, and full way back to start the next 120° rotation which is N-rich region. The extra B (or N) atoms are located at B-B (or N-N) bonds, which cannot be eliminated by local reconstruction of 5|7s into 4|8s because there are no compensation N-N (or B-B) bonds along the same GB. On the other hand, in a global view, extra B (N) from B-rich (N-rich) GB could probably pair up with each other and result in the elimination of polarity. Two examples of GBs 'reactions' are demonstrated in Figure 4, where polarity disappears through 5|7s → 4|8s reconstruction. We find that these reactions are endothermic, illustrated by the energies of 'products' (green squares) and 'reactants' (green circles) in Figure 3. The dashed lines connect the 'average' energies of two GBs with reverse elemental polarity, *i.e.* one is B-rich (with angle $α$) and the other must be N-rich (with angle 120°+$α$). Increasing chemical potential of B decreases the energy of B-rich GB, but increases that of N-rich one. Nevertheless, the average value keeps unaffected by chemical potential. The averaged value can also be used to quantify the energies of GBs in multiple BN layers, which follow A-A' stacking (B on top of N or vice versa)[26] thus might guide the B-rich and N-rich GBs stacked together.

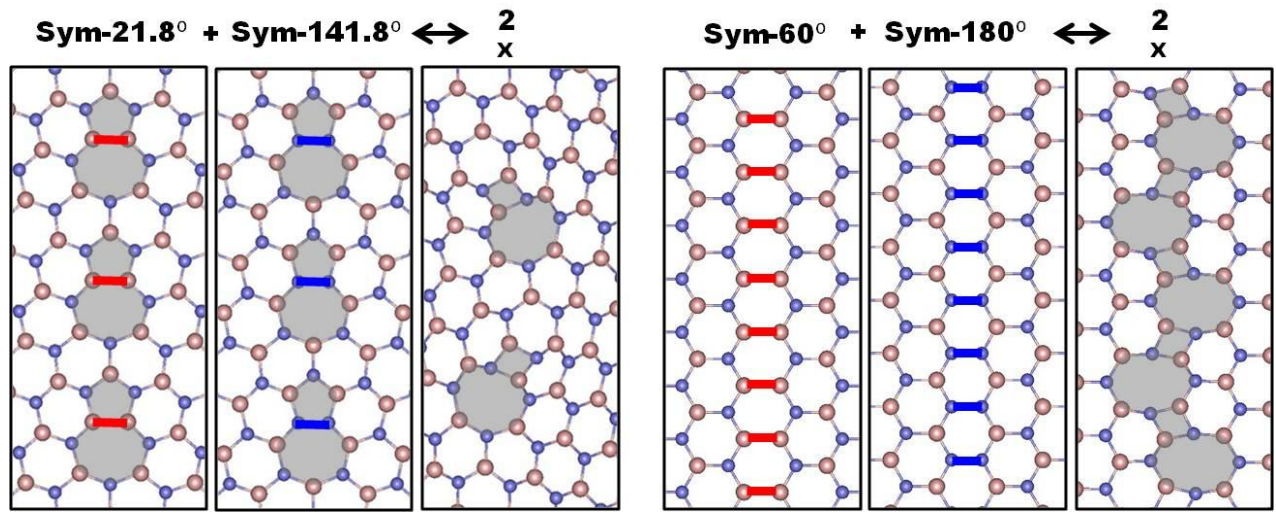

**Figure 4**. Transformation "reactions" of polar GBs. Extra B and N atoms from GBs pair up and 4|8s can be generated during these reactions.

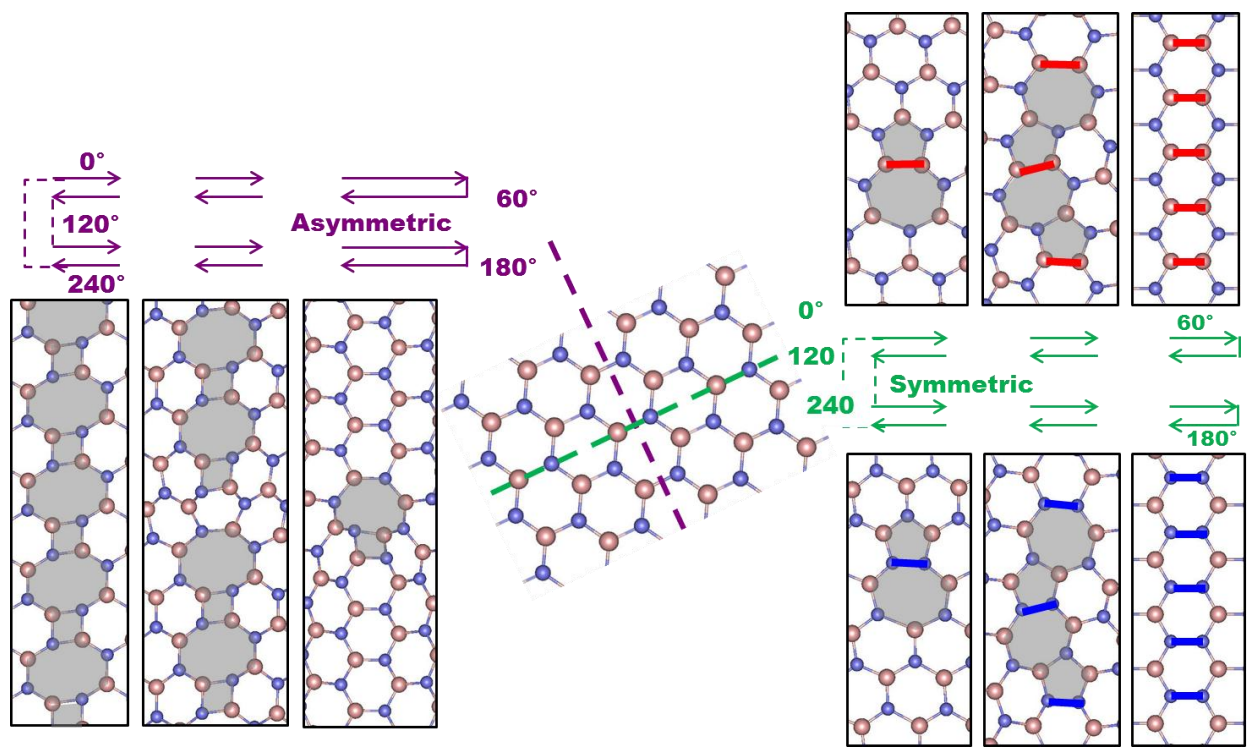

**Figure 5**. Ground state structures of GBs as a function of tilt angle. They are similar to those shown in Figure 2 except that asymmetric GBs are nonpolar and made by 4|8s.

Figure 5 summarizes the ground state structures of GBs. The asym-GBs are un-polar and constituted by 4|8s. The sym-GBs are polar and composed by 5|7s. All of them can co-exist in polycrystalline BN, suggesting a wide variation of structures and related properties under different circumstances. The balance between substrate flattening and defects induced corrugation leads to three-dimensional landscape in BN membrane, with its roughness scaling inversely with the defects concentration, similar to discussed for graphene.[7] Similar to graphene, GBs in BN should serve as fracture nucleation center under high tension load,[18,19] while their mechanic properties are more complicated due to the existence of homo-elemental bonds and the variation between 5|7s and 4|8s. For example, the better accommodation of 5|7 to the elastic strain suggests possible transformation from 4|8 to 5|7 under mechanical loading. This topic deserves further investigation.

More interestingly, since B-N bond has ionic nature with electrons transferred from B to N, the polar GBs carry net charges brought by extra atoms. Figure 6b plots the computed electrostatic potential from sym-60° (B-rich) to sym-180° (N-rich) GB (the potential is averaged along GB direction in the plane 3 Å above 2D BN). The monotonic decrease of potential indicates B-rich GB is positively charged and N-rich GB carries negative charges, agreeing with the charge transfer in perfect BN. This intrinsic electric field, which does not exist in graphene, could be utilized to detect GBs using Kelvin probe force microscope or electrostatic force microscope. Furthermore, the electronic properties of GBs are also remarkable. The electronic signals should be most prominent at GB with highest concentration of defects, for example, boundaries full of B-B bonds (sym-60°), of N-N bonds (sym-180°), or 4|8s (asym-0°). Their density of states (DOS) are shown in Figure 6a. Sym-180° have occupied states contributed by extra N atoms appearing above the valence band maximum (VBM) of perfect BN, while for

sym-60°, extra B atoms bring in unoccupied states below the conduction band minimum (CBM). Analysis of energy-decomposed charge density plot indicates that these extra electronic states are located at GBs only, and decrease the bandgap by 38% (1.5eV/4eV, although DFT cannot give the precise value of bandgap), in contrast to GBs in graphene which generally open the bandgap.[14,16] However, 4|8s do not shift VBM and CBM significantly. The decreased bandgap could probably create optical peaks closer to the visible light region and might be also utilized to identify the GBs. To facilitate experimental observations, simulated scanning tunneling microscope (STM) images[27] are also shown. They are conducted at the constant height model, with 2 Å above BN and positive bias 0.5 Volt with respect to VBM. Distinct electronic fingerprints can be clearly seen through these images. B-rich sym-60° GB has higher charge density at B-B bonds, while its N-rich analogy sym-180° GB, has intense charge at N atoms which form N-N bonds. 4|8s induce charge accumulation around N atoms as well, in spite of isoelectronic structure as perfect BN.

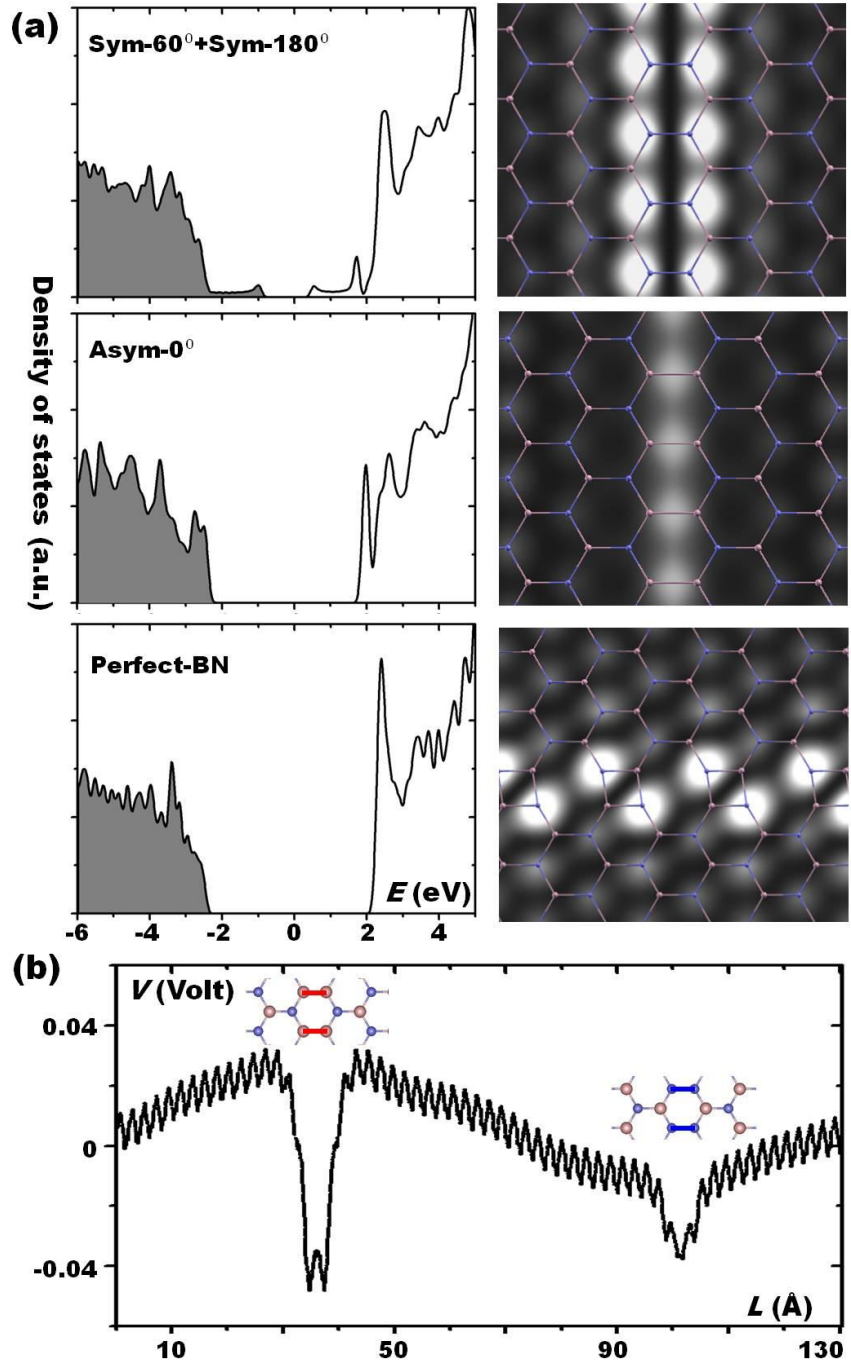

**Figure 6**. (a) Left: density of electronic states (left) for perfect BN (bottom), BN containing asym-0° GB (middle), and BN containing sym-180° and sym-60° GB (top). Shadowed areas indicate the occupied states. Right: simulated STM image for sym-180° (top), sym-60° (middle) and asym-0° (bottom) GBs. Atomic structures are put in front of STM images. Red balls

represent B and blue for N. (b) Electrostatic potential distribution in the plane 3 Å above the polycrystalline 2D BN which contains sym-60° and sym-180° GBs. The potential is averaged along GB directions.

**CONCLUSION:**

To summarize, we present the topology and energies of dislocations and GBs in 2D h-BN sheets. Different from graphene, dislocation with burgers vector (1,1) is comprised from 4|8, instead of 5|7s. In fact, the similar alteration has been observed in BN fullerene cages and BN-tube caps, where even-membered rings exist instead of odd-membered rings, as a distinguishing feature different from carbon materials.[23,28,29] Depending on the tilt angle of grains, GBs can be either polar (B-rich or N-rich), constituted by 5|7s, or nonpolar, composed of 4|8s. Polar GBs possess net charges and smaller bandgap compared to perfect BN. Given the hetero-elemental nature of many other 2D systems, like transition-metal disulfides, the results could probably be generalized into these materials, though detailed energy calculations are desired. The extra variable in atom type, would open many possibilities for structures of defects, as well as expected novel properties and applications.

**METHODS:**

Density functional theory (DFT) calculations are performed with Perdew-Burke-Ernzerhof parameterization (PBE)[30] of generalized gradient approximation (GGA) and projector-augmented wave (PAW) potentials,[31,32] as implemented in Vienna Ab-initio Simulation Package (VASP). [33,34] All structures are fully relaxed until the force on each atom is less than 0.01 eV/ Å.

Vacuum spacing is kept larger than 10 Å to keep the spurious interaction negligible. To compute formation energy, non-hexagon polygons are embedded in a hexagonal sheet with their distance to the edge larger than 15 Å. GBs are embedded in ribbons with width larger than 20 Å. Sufficient Monkhorst-Pack k-points[35] are used along periodic directions to ensure energy convergence.

**ACKNOWLEDGEMENTS:**

This work was supported by the Department of Energy, BES Grant No. ER46598, and the US Army Research Office MURI Grant No. W911NF-11-1-0362. The computations were performed at (1) the National Institute for Computational Sciences, through allocation TG-DMR100029; (2) the National Energy Research Scientific Computing Center, which is supported by the Office of Science of the U.S. Department of Energy under Contract No. DE-AC02-05CH11231; and (3) the Data Analysis and Visualization Cyberinfrastructure funded by NSF under grant OCI-0959097. Y.L. thanks Mr. Luis Flores for proofreading.

**TOC**

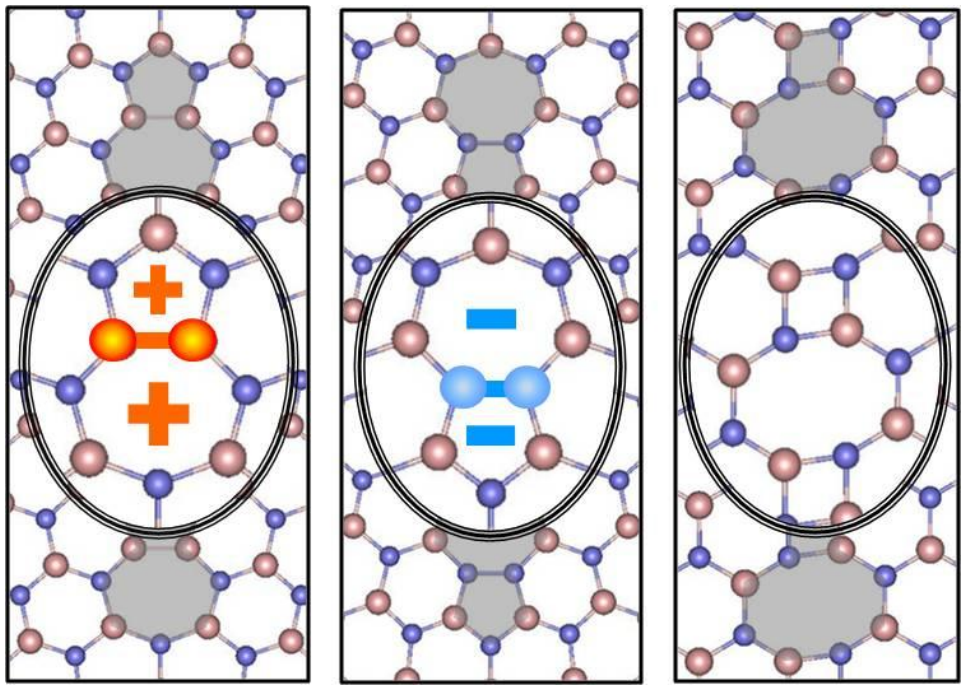